\documentclass[a4paper,10pt,]{article}

\usepackage{amsmath}
\usepackage{amssymb}
\usepackage{mathtools}
\usepackage{graphicx}
\usepackage{latexsym}
\usepackage{color}
\usepackage[colorlinks=true,linkcolor=red]{hyperref}%

\usepackage{authblk}

\usepackage{mathabx}

\newcommand{\fie}{\varphi}

\newcommand{\ve}[2]{\begin{pmatrix} #1 \\  #2 \end{pmatrix}}

\newcommand\changes[1]{\textcolor{black}{#1}}

\begin{document}

\title{On the singularity structure of the discrete KdV equation}
\author[1]{Doyong Um
}
\author[2]{Ralph Willox\thanks{willox@ms.u-tokyo.ac.jp}}
\author[3]{Basil Grammaticos}
\author[3]{Alfred Ramani}
\affil[1]{Environmental Sciences (PEAK), College of Arts and Sciences, the University of Tokyo, 3-8-1 Komaba, Meguro-ku, 153-8914 Tokyo, Japan}
\affil[2]{Graduate School of Mathematical Sciences, the University of Tokyo, 3-8-1 Komaba, Meguro-ku, 153-8914 Tokyo, Japan}
\affil[3]{ IMNC, Universit\'{e} Paris VII \& XI, CNRS, UMR 8165, Orsay, France}

\maketitle

\begin{abstract}
The discrete KdV (dKdV) equation, the pinnacle of discrete integrability, is often thought to possess the singularity confinement property because it confines on an elementary quadrilateral. Here we investigate the singularity structure of the dKdV equation through reductions of the equation, obtained for initial conditions on a staircase with height 1, and show that it is much more subtle than one might assume. We first study the singularities for the mappings obtained after reduction and contrast these with the singularities that arise in non-integrable generalizations of those mappings. We then show that the so-called `express method' for obtaining dynamical degrees for second order mappings can be succesfully applied to all the higher order mappings we derived. Finally, we use the information obtained on the singularity structure of the reductions to describe an important subset of singularity patterns for the dKdV equation and we present an example of a non-confining pattern and explain why its existence does not contradict the integrability of the dKdV equation.
\end{abstract}

\section{Introduction}\label{intro}
The Korteweg-de Vries (KdV) equation is widely regarded as the integrable system {\sl par excellence}, not least because the modern era of integrability is intimately related to the study of its properties. When setting out to analyze the Fermi-Pasta-Ulam-Tsingou (FPUT) problem \cite{FPUT}, Kruskal and Zabusky remarked that the continuum limit of the system in question was the KdV equation \cite{kruskal}. The FPUT system could therefore be considered to be a discretisation of KdV and while studying the properties of this discretisation, they remarked that the solitary waves (the existence of which was already established by Boussinesq \cite{boussinesq} and then, later, rediscovered by Korteweg and De Vries \cite{KDV}) were interacting in an elastic way. The soliton was discovered and the rest is history.

A discrete analogue of the KdV equation, which preserves the integrability properties  of the latter, was proposed by Hirota \cite{hirotadKdV} soon after the integrability of the continuous KdV equation was first established. In fact, Hirota had singlehandedly made huge inroads into the domain of integrable discrete systems, already years before the integrability community became interested in them. Hirota's studies culminated in the discovery of the discrete form of the Kadomtsev-Petviashvili (KP) equation, of which the discrete KdV is a two-dimensional reduction.
The discrete KdV \changes{(dKdV)} equation, as derived by Hirota, has the form (in  current, modern, notation)
\begin{equation}
x_{m+1,n+1}=x_{m,n}+{1\over x_{m+1,n}}-{1\over x_{m,n+1}}.\label{dKdV}
\end{equation}
Hirota proved the existence of an arbitrary number of soliton-like solutions for \eqref{dKdV} and constructed a B\"acklund transformation and Lax pair for this equation. What was missing was the proof that the KdV equation satisfies a discrete integrability criterion which could be considered a discrete analogue of the Painlev\'e property for differential systems \cite{Painleve}.  This had to wait for several more years.

While studying properties of discrete systems two of  the present authors, in collaboration with Papageorgiou, made a remarkable observation \cite{KdVsing}. Discrete systems which could be considered as integrable (according to other criteria) had a special singularity structure. In particular, any singularity appearing spontaneously, due to a particular choice of initial conditions (a situation which in the continuous case was referred to as a `movable' singularity) disappeared after a certain number of iteration steps. This property came to be known as {\sl singularity confinement} and the current conjecture is that all discrete systems integrable through spectral methods do possess this property. 

It is not astonishing that the system for which this property was first observed was precisely the discrete KdV equation. That first study was limited to initial conditions given on a staircase of height 1 and width 1 but as we shall argue in this paper, it is imperative to study also other types of initial conditions in order to truly understand the singularity properties of the equation.

\begin{figure}[h]
\begin{center}
\resizebox{5.15cm}{!}{\includegraphics{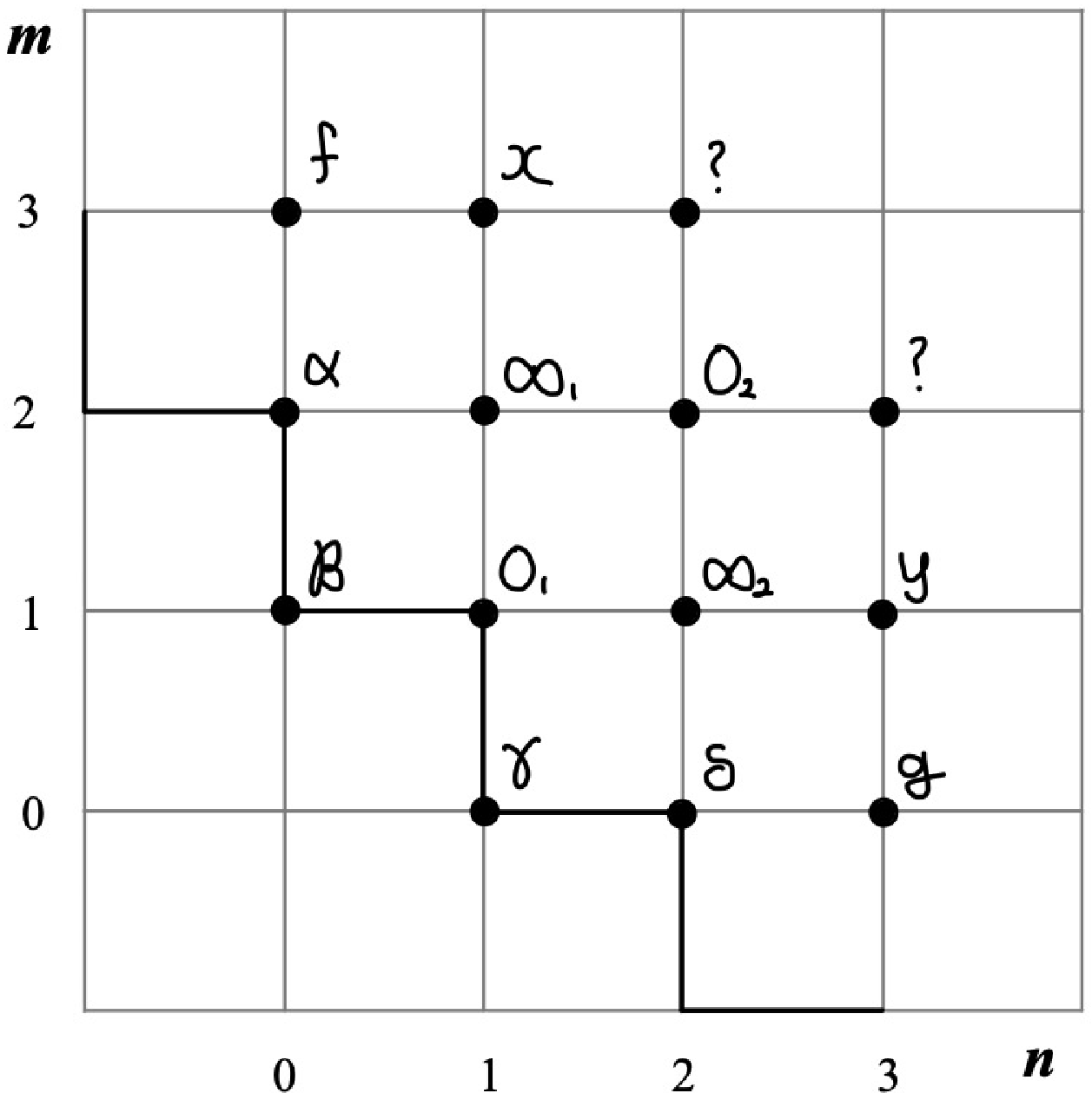}}
\caption{The singularity pattern for the lattice KdV equation for $(1,1)$ staircase.}
{\small The numbers appearing as subscripts in $0$ and $\infty$ are used to indicate their positions in the lattice.}
\label{Fig2}
\end{center}
\end{figure}

For the benefit of the reader though, let us first explain in detail the case of initial conditions on a 1-1 (height 1, width 1) staircase, the simplest possible situation, for a slightly generalized KdV-type equation:
\begin{equation}
x_{m+1,n+1}=x_{m,n}+{a\over x_{m+1,n}}-{b\over x_{m,n+1}},\label{gendKdV}
\end{equation}
with $a,b\in\mathbb{C}\setminus\{0\}$, which coincides with \eqref{dKdV} when $a=b$ (whereupon the constant can be put to 1 by rescaling $x$). As mentioned above, we shall consider the case of initial data (in $\mathbb{C}$) given on a 1-1 staircase (the black line in Figure \ref{Fig2}), from which the solution to \eqref{gendKdV} evolves (with values in $\mathbb{C}\cup\{\infty\}$) by iteration in the northeast direction in the $(n,m)$-plane.

A problem in this evolution can arise when a 0 appears ``spontaneously'' at some lattice site, due to some particular choice of initial conditions, which given the form of the equation is perfectly possible. However, the effect of such a zero is determined by the values of $x_{m,n}$ at neighbouring lattice sites, which is why we model this situation by assuming, without loss of generality, that we have in fact  a 0 value at a site {\sl on} the initial 1-1 staircase. Let us also assume that there is only one such initial value. Problems arise if this 0 is located at a northeast (i.e. protruding) corner of the staircase, as depicted in Figure \ref{Fig2}. Starting from initial values $x_{2,0}=\alpha,x_{1,0}=\beta,x_{1,1}=0,x_{0,1}=\gamma$ and $x_{0,2}=\delta$ we obtain $x_{2,1}=\infty$, $x_{1,2}=\infty$, $x_{2,2}=0$, and arrive at indeterminate results of the type $\infty - \infty$ for $x_{3,2}$ and $x_{2,3}$. To lift this indeterminacy, we assume that $x_{1,1}=\epsilon$ and perform the above iteration for this initial condition. As we shall see, when $a=b$, taking the limit for $\epsilon \rightarrow 0$ we find that $x_{3,2}$ and $x_{2,3}$ take regular values (i.e., generically nonzero finite values). Thus, we say that this singularity is {\em confined} for the dKdV equation \eqref{dKdV}, and the corresponding {\em singularity pattern} is as depicted in Figure \ref{Fig2} \cite{KdVsing}.

\vspace{2mm}

The detailed singularity analysis for the initial value $x_{1,1}=\epsilon \;(=0_1)$ for generic values for $a$ and $b$, yields
$$
x_{2,1}=\beta+\frac{a}{\alpha}-\frac{b}{\epsilon}\approx-b\epsilon^{-1} \ (=\infty_{1}), \quad x_{1,2}=\gamma+\frac{a}{\epsilon}-\frac{b}{\delta}\approx a\epsilon^{-1}\ (=\infty_{2}),
$$
and
$$
x_{2,2}=\epsilon+\frac{a}{\infty_{1}}-\frac{b}{\infty_{2}}=(1-\frac{a}{b}-\frac{b}{a}) \epsilon+{o}(\epsilon) \ (=0_{2}).
$$
Since the initial values $x_{3,-1}$, $x_{2,-1}$ and $x_{2,0}$ are all regular values, we obtain a regular (generic) value at $x_{3,0}$. Similarly, we obtain a regular value at $x_{0,3}$. Let us denote these as $x_{3,0}=f$ and $x_{0,3}=g$. We then also have finite values at the subsequent iterations:
$$
\begin{aligned}
    x_{3,1} &= \alpha+\frac{a}{f}-\frac{b}{\infty_{1}}=\alpha+\frac{a}{f}+\epsilon+ o(\epsilon)\approx \alpha+\frac{a}{f} \ (=x), \\
    x_{1,3} &= \delta+\frac{a}{\infty_{2}}-\frac{b}{g}=\delta-\frac{b}{g}+\epsilon+ o(\epsilon)\approx\delta-\frac{b}{g} \ (=y).
\end{aligned}
$$
Now, if $a=b$, these resolve the ambiguities at the next step:
$$
\begin{aligned}
    x_{3,2} &= \infty_{1} + \frac{a}{x} - \frac{b}{0_{2}} = -\frac{b(a-b)^2 \epsilon^{-1}}{a^2-ab+b^2}+\left(\gamma - \frac{a}{\delta}+\frac{a}{\alpha+a/f}\right) +{\cal O}\big( (a-b),\epsilon\big)\approx \gamma - \frac{a}{\delta}+\frac{a}{\alpha+a/f}, \\
    x_{2,3} &= \infty_{2} + \frac{a}{0_{2}} - \frac{b}{y} = \frac{a(a-b)^2 \epsilon^{-1}}{a^2-ab+b^2}+\left(\beta + \frac{a}{\alpha} - \frac{a}{\delta - a/g}\right) +{\cal O}\big( (a-b),\epsilon\big)\approx\beta + \frac{a}{\alpha} - \frac{a}{\delta - a/g} .
\end{aligned}
$$
Note that the above cancellations of infinities cannot happen if $a\neq b$, and in that case infinities occur at $x_{3,2}$ and $x_{2,3}$ which then persist throughout, in the northeast direction, and the singularity at $x_{1,1}=0$ then no longer confines in a finite number of iteration steps. This implies that requiring singularity confinement  (the {\sl singularity confinement criterion}) already for this simplest singularity, actually suffices to distinguish between the integrable and non-integrable case of the  generalised dKdV equation \eqref{gendKdV}. Hence the usefulness of the criterion. \changes{Note that had we assumed $a$ and $b$ to be functions of $m$ and $n$, the study of this singularity would have sufficed to fix the precise  $m,n$ dependence of $a$ and $b$ that is required for integrability \cite{fulldeauto}. (See also \cite{kajiohta} for a detailed account of the properties of the resulting {\sl non-autonomous} dKdV equation.) }

In light of this result one could be tempted to immediately conclude that the dKdV equation satisfies the singularity confinement criterion, but this would be too hasty as the above singularity is of course not the only possible one. It is easy to verify that a zero in the southeast (sunken) corner of a 1-1 staircase is never singular. One can, however, imagine situations where two or more zeros appear at adjacent protruding corners of the staircase, \changes{as already investigated in the original paper on singularity confinement \cite{KdVsing} (see e.g. figure 2 in that reference) where it was shown that the ensuing singularities indeed disappear, albeit after more iteration steps.}

\begin{figure}[h]
\begin{center}
\resizebox{13.5cm}{!}{\includegraphics{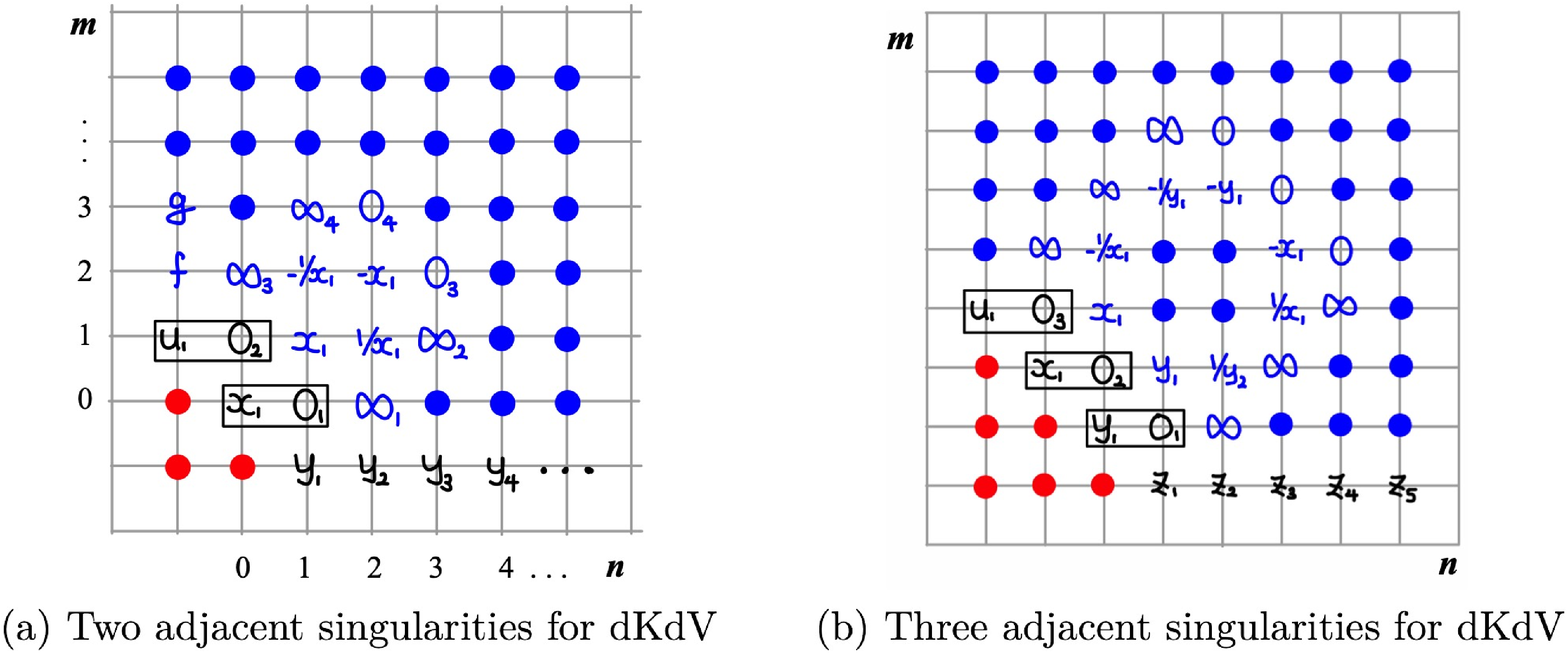}}    
\caption{Initial 1-1 staircases with a higher number of singularities.}
\label{Caseopop}
\end{center}
\end{figure}

\changes{Here we are interested in the special case where these zeros are exactly identical. Figure \ref{Caseopop}(a) shows the singularity pattern obtained for identical zeros on two adjacent protruding corners of a 1-1 staircase. With a higher number of adjacent (identical) zeros on such a staircase the singularity pattern becomes more and more extended, as shown in Figure \ref{Caseopop}(b).}
While the details of the singularity analysis soon become unwieldy it is easy to convince oneself that all such singularity patterns are rhombus shaped, with parallel lines of $n$ zeros and $n$ infinities. This means that each such singularity pattern is confined in a finite number of iterations, for a finite number of zeros on the initial 1-1 staircase. 

\begin{figure}[t]
\begin{center}
\resizebox{12.5cm}{!}{\includegraphics{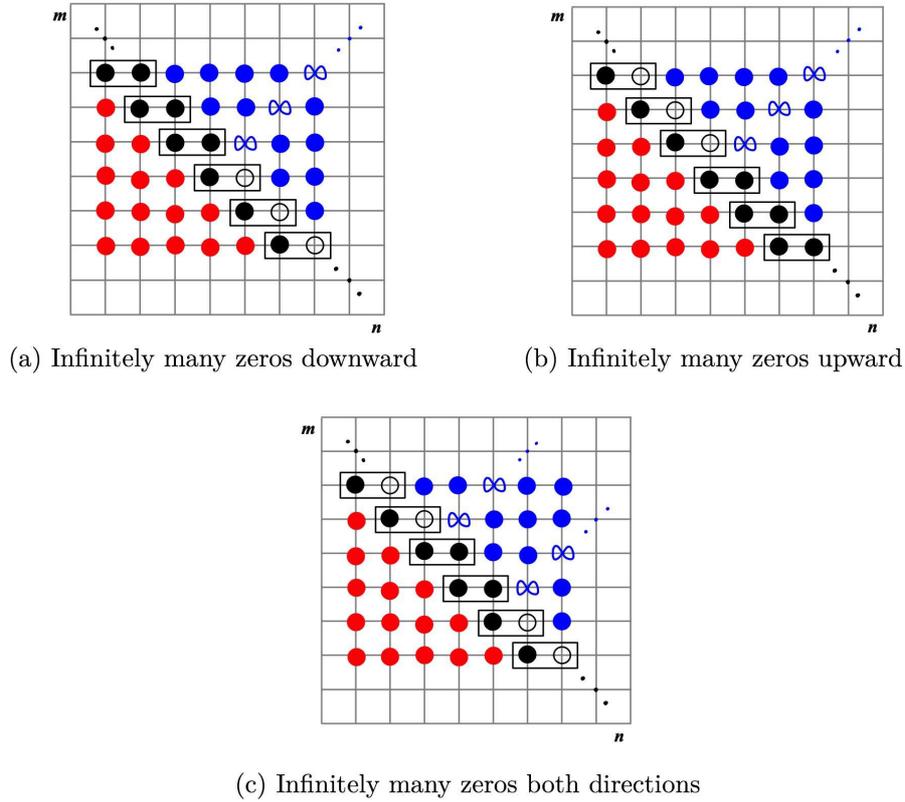}}
\caption{Unconfined lattice singularity patterns for dKdV, starting from infinitely many zeros.}
\label{p1q1inf}
\end{center}
\end{figure}

It is important to consider what happens when the number of \changes{such} singularities on the initial staircase is infinite. When $x_{m_0-k,n_0+k}=0$ for all $k\in\mathbb{Z}$ (for some choice of $(m_0,n_0)$) no singularities arise at all as long as all other (non-zero values) on the initial staircase are different -- we shall come back to this point, later on, in section \ref{KdVsing}. However, one can also set infinitely many (different) generic values \changes{alongside} infinitely many zeros at protruding corners of the initial staircase, by choosing for instance $x_{m_0-k,n_0+k}=0$ for $k\in\mathbb{Z}_{\geq0}~ \text{or} ~\mathbb{Z}_{\leq0}$. The corresponding singularity patterns are shown in Figures \ref{p1q1inf} (a) and (b). For these initial conditions, a zero at the boundary produces an infinity just above it (in (a)) or to the right of it (in (b)). This infinity then spreads out indefinitely along the diagonal $m=n$ and no rhombus-like confined singularity pattern is obtained. A singularity pattern as in Figure \ref{p1q1inf} (c) is also possible. Clearly, having infinitely many zeros among the initial conditions can and does give rise to {\sl unconfined} singularities for dKdV, its integrability notwithstanding. (A similar observation can be made for the 2D-discrete Toda equation \cite{gromov, tsuboi}.)

All the arguments presented above were based on a specific type of initial condition, namely a height 1--width 1 staircase. However, infinitely many other possibilities do exist. \changes{As pointed out in \cite{atkinson1} where the possible singularities in type-Q equations in the ABS-classification \cite{ABS} are investigated from the point of view of multidimensional consistency, the singular values that may appear in non type-Q equations like the dKdV equation  present particular difficulties (see also \cite{atkinson2}).}
While an exhaustive study of all \changes{possible singularity patterns} looks like a prohibitively difficult task, one can gain substantial insight into the workings of singularity confinement by considering the approach introduced by one of us in collaboration with Ablowitz and Segur in the continuous setting \cite{ars}. That approach, often referred to as the ARS method, is based on the elementary observation that if a 2-D system is integrable then its 1-D reductions should possess the same property. The present paper is directly inspired by this. Instead of tackling the problem of the singularity structure of the solutions of dKdV, it considers the family of reductions of the type $x_{m+1,n}=x_{m,n+q}$ ($q\in\mathbb{Z}_{\geq1}$) and studies their singularities. More precisely, we will show that for general $q\in\mathbb{Z}_{\geq2}$, the mapping of order $q+1$ that is obtained from the reduction has exactly two singularities which need to be discussed and we investigate the confinement properties of these singularities for all mappings obtained from the dKdV equation \eqref{dKdV} as well as from the non-integrable case ($a\neq b$) of equation \eqref{gendKdV}. We then go on to show that the so-called {\sl express method} we introduced in \cite{express} for determining the dynamical degree of second order mappings, can in fact also be successfully applied to all the mappings we obtain as reductions of equations \eqref{dKdV} and \eqref{gendKdV}. Finally, in section \ref{KdVsing}, we investigate what, if anything, the singularities for the mappings obtained from the reduction can tell us about singularities that may arise for the dKdV equation. Although this investigation cannot be exhaustive, it does allow us to identify an important case in which a single singular value gives rise to a non-confining singularity pattern for dKdV. We also explain why the existence of such a pattern does not contradict the integrabilty of the equation. 

\section{A family of reduced mappings, their singularities and degree growth}\label{redmapp1}
\begin{figure}[h]\vspace{2mm}
\begin{center}
\resizebox{11.5cm}{!}{\includegraphics{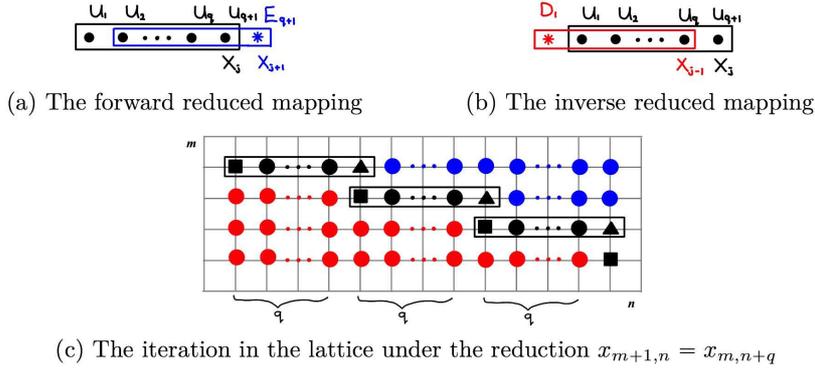}}
\caption{Identification of initial conditions for the reduced mapping \eqref{map1p} and dKdV.}
\label{Redp1}
\end{center}
\end{figure}

The reduction of \eqref{gendKdV} for generic $a, b\in\mathbb{C}\setminus\{0\}$, under the constraint $x_{m+1,n}=x_{m,n+q}$ ($q\in\mathbb{Z}_{\geq1}$), gives rise to the following \changes{$q+2$} point mapping,
\begin{equation}
x_{m,n} = x_{m,n-q-1} + \frac{a}{x_{m,n-1}} - \frac{b}{x_{m,n-q}},\label{equ1p}
\end{equation}
where the subscript $m$ is now superfluous as it is common to all terms. We represent this mapping as a birational mapping on the $q+1$-dimensional projective space $\left(\mathbb{P}^1\right)^{q+1}:=\mathbb{P}^{1} \times \cdots \times \mathbb{P}^{1}$, by defining new variables
\begin{equation}
u_{j} := x_{m,j-1} \quad  (j=1,\cdots,q+1).\label{defu}
\end{equation}
We can then rewrite \eqref{equ1p} as the birational mapping 
\begin{equation}
\fie_q : \left(\mathbb{P}^1\right)^{q+1} \dashrightarrow \left(\mathbb{P}^1\right)^{q+1}\,,\quad 
\begin{pmatrix}{u_{1}}\\ \vdots\\ u_{q}\\{u_{q+1}}\end{pmatrix}  \mapsto 
\begin{pmatrix} {u_{2}}\\ \vdots \\u_{q+1} \\ E \end{pmatrix}
\quad\text{where}\quad E := {u_1}+\frac{a}{u_{q+1}}-\frac{b}{u_2}.
\label{map1p}
\end{equation}
Its inverse mapping is given by
\begin{equation}
\fie_q^{-1} : \left(\mathbb{P}^1\right)^{q+1} \dashrightarrow \left(\mathbb{P}^1\right)^{q+1}\,,\quad 
\begin{pmatrix}{u_{1}}\\ \vdots\\ u_{q}\\{u_{q+1}}\end{pmatrix}  \mapsto 
\begin{pmatrix} D \\ u_{1} \\ \vdots \\u_{q} \end{pmatrix}
\quad\text{where}\quad D:= u_{q+1}+\frac{b}{u_{1}}-\frac{a}{u_{q}}.
\label{invmap1p}
\end{equation}
If we take the values $u_1=x_{m,0}, u_2 = x_{m,1},\hdots,$ $u_{q+1}=x_{m,q}$ as initial conditions for the mapping \eqref{map1p}, we have a straightforward 1:1 correspondence between the forward (and backward) iterations of $\fie_q$, depicted in Figure \ref{Redp1} (a) (and (b)), and the foreward evolution in the northeast direction (blue dots) or the backward evolution (red dots) of the lattice equation \eqref{gendKdV}, for periodic initial conditions $x_{m-k,qk}=u_1, x_{m-k,qk+1}= u_2,\hdots,x_{m-k,qk+q}=u_{q+1}$ ($^\forall k\in\mathbb{Z}$) on a staircase of height 1 and width $q$ (as depicted in Figure \ref{Redp1} (c)). This will be important later on.

The case $q=1$ is best treated separately as it is almost trivial. Indeed, for $q=1$ we have the birational mapping
\begin{equation}
\fie_1 : \mathbb{P}^1\times \mathbb{P}^1 \dashrightarrow \mathbb{P}^1\times \mathbb{P}^1\,,\quad 
\begin{pmatrix}{u_{1}}\\ {u_{2}}\end{pmatrix}  \mapsto 
\begin{pmatrix} {u_{2}}\\ u_1 + \frac{a-b}{u_2} \end{pmatrix},
\label{map1p1q}
\end{equation}
which is obviously 2-periodic when $a=b$. Moreover, since $\fie_1$ leaves the rational fibration $u_1 u_2 = \kappa$ ($\kappa\in\mathbb{P}^1$) invariant, it can be linearized and therefore solved explicitly even when $a\neq b$. The general solution (for $n\in\mathbb{Z}_{\geq0}$) can be written as 
$$\fie_1^n\begin{pmatrix}{u_{1}}\\ {u_{2}}\end{pmatrix} = \begin{pmatrix}{\omega_{n}}\\[-2mm]\\ \dfrac{u_1 u_2 + n\gamma}{\omega_{n}}\end{pmatrix},\qquad \text{where}\quad  \omega_n =
\begin{dcases}
    u_2 \prod_{k=1}^{\ell}\dfrac{u_1 u_2+2k\gamma}{u_1 u_2+(2k-1)\gamma}, \quad \text{if $n=2\ell+1$} \\
   u_1 \prod_{k=0}^{\ell-1} \dfrac{u_1 u_2+(2k+1)\gamma}{u_1 u_2+2k\gamma}, \quad \text{if $n=2\ell$} ,
\end{dcases}
 $$
for $\ell\in\mathbb{Z}_{\geq0}$ and $\gamma=a-b$. When $a=b$, the mapping $\fie_1$ is linear and has no singularities. When $a\neq b$, $\fie_1$ has exactly one  singularity, the curve $u_2=0$ in $\mathbb{P}^1\times \mathbb{P}^1$ which collapses to a point. This singularity is unconfined (in accordance with the general classification result on second order birational mappings in \cite{dillerfavre} which says that a second order mapping with an invariant rational fibration should have such a singularity) and corresponds to the {\sl unconfined} singularity pattern
\begin{equation}
\left\{\ve{u_1}{0},\ve{0}{\infty},\ve{\infty}{0},\ve{0}{\infty},\cdots \right\}.\label{patt1q}
\end{equation}
This result is obtained by iterating the mapping for an initial condition where $u_1$ is a generic complex number and where $u_2=\epsilon$, and by then taking the limit $\epsilon\to0$  just as for the dKdV equation in the introduction. From this calculation it is readily seen that although the point $^t\!\left(\infty~\,0\right)$ is an indeterminate point for the mapping $\fie_1$ when $a\neq b$, 
the information on $u_1$ that disappeared in the singularity is never recovered at the limit $\epsilon\to0$, i.e. the indeterminacy is never `lifted'. Moreover, it is also easily seen that the backward orbit of the curve $u_2=0$ through $\fie_1$ consists only of curves, i.e. it does not lead to any singularities. Hence we can conclude that the singularity pattern \eqref{patt1q} is indeed unconfined (in the sense of \cite{unconfined}) and not {\sl anticonfined} \cite{anticonf, review}. 

When $q>1$ we obtain higher order mappings for which there are no general classification results  as in \cite{dillerfavre} and for which the relationship between the singularity structure and integrability properties of the mapping remains to be clarified. The guiding principle in our singularity analysis for such mappings will be to focus on singular varieties of co-dimension 1 in $\left(\mathbb{P}^1\right)^{q+1}$ and to ignore all singularities with greater co-dimension. The idea to restrict our definition of a singularity to co-dimension 1 varieties is based on the assumption that just as in the case of second order mappings, where integrable (non-linearisable) birational mappings can be regularised by a finite number of blow ups to an automorphism on a rational surface \cite{sakai,takenawaHV}, integrability for higher dimensional birational mappings must be intimately related to the ability to blow up the mapping to a pseudo-isomorphism between blow-up spaces \cite{carstea}. Such a pseudo-isomorphism is only required to be regular on co-dimension 1 subvarieties and can still have singular behaviour on varieties with higher co-dimension \cite{bedford}.

Let us therefore define the subspace $X_{0} \subset \left(\mathbb{P}^1\right)^{q+1}$ of co-dimension 1:
\begin{equation}
X_{0}=\left\{\mathbf{x}_{0}=(u_1,\cdots,u_{q+1}) \in \left(\mathbb{P}^1\right)^{q+1} \big| \,^{\exists!} k \in \mathbb{Z}, 1 \leq k \leq q+1~\,s.t.~ u_k=0 \right\}.\label{X0}
\end{equation}
Given that any generic vector $\mathbf{x}_{0}\in X_0$ can be taken as a representative for $X_0$, we may interpret \eqref{map1p} as mapping the subspace $X_0$ to a new subspace $X_{1}$ and we can then, recursively, define the images $X_j$ of $X_0$ under the composition of $\fie_q$ by 
$$ 
X_{0}:=\fie_q^{0}(X_{0}) \quad X_{n}:=\fie_q^{n}(X_{0}) = 
\begin{cases}
\begin{aligned}
(\fie_q \circ \fie_q^{n-1})(X_{0}), \quad & (n \in \mathbb{Z}_{>0}) \\
(\fie_q^{-1} \circ (\fie_q^{-1})^{|n|-1})(X_{0}), \quad & (n \in \mathbb{Z}_{<0}).
\end{aligned}
\end{cases}
$$
We shall denote vectors in $X_j$ by $\mathbf{x}_{j}$ $(j \in \mathbb{Z})$ throughout this paper. 

In the case of second-order rational mappings the notion of a singularity refers to a loss of degree of freedom at some iteration: i.e. a mapping is singular at $x_n$ if the value of $x_{n+1}$ is independent of that of $x_{n-1}$. Similarly, for a \changes{higher order} mapping, we shall say that a singularity occurs at $X_n$ if $X_{n+1}$ has a higher co-dimension than $X_{n}$. (On the contrary, if  the co-dimension of $X_{n+1}$ is lower than that of $X_n$, then we say that $\fie_q$ is indeterminate on $X_n$.)
Since we are dealing with \changes{$q+2$} point mappings,  for which the first $q$ entries in $\fie_q(\mathbf{x}_{j})\in X_{j+1}$ are the same as the last $q$ entries in $\mathbf{x}_{j}\in X_{j}$, it is easily verified that the general mapping \eqref{map1p}, for $q\geq2$, has exactly two singularities with co-dimension 1: $X_{0}$ with $u_2=0$ or $u_{q+1}=0$. 

In the spirit of the ARS approach, since the generalized dKdV equation \eqref{gendKdV} is known to be integrable when $a=b$, we know that all its reductions in that case must yield integrable mappings, the singularities of which must have properties compatible with integrability. \changes{In fact, it has been shown in \cite{honeetal} that the mapping \eqref{map1p} for $a=b$ is Liouville integrable for general $q\geq3$.} On the other hand, when $a\neq b$ the equation is thought to be non-integrable and we expect its reductions to yield (at least some) non-integrable mappings, a property which should also be reflected in the nature of their singularities. In fact, we shall show that all reductions for $q\geq 2$ in the case $a\neq b$ are non-integrable. For this, we first analyze the properties of the singularities of $\fie_q$ on $X_{0}$ with $u_2=0$ or $u_{q+1}=0$ for the integrable case ($a=b$) and for the non-integrable case ($a\neq b$),  separately.

\subsection{The integrable case $(a=b)$ of mapping \eqref{map1p} for $q\geq 2$}\label{singpataeqb}
As mentioned before, if $a=b$  the parameter dependence in equation \eqref{equ1p} can be removed by scaling $x_{m,n} \mapsto \sqrt{a} x_{m,n}$ and since the singularity confinement property is invariant under rescaling of the dependent variables, we may assume $a=b=1$.

Let us first analyse the singularity patterns for the integrable reduced mapping \eqref{map1p} for $q=2$. As explained above, this mapping  has just two singularities: $X_{0}$ with $u_{2}=0$ and $X_{0}$ with $u_3=0$. Both singularities are confined, but they correspond to singularity patterns of different types. The first singularity, at $X_{0}$ with $u_{2}=0$, corresponds to a (confined) singularity pattern which repeats with period 4:
\begin{equation}
\Bigg\{ \begin{pmatrix} u_{1} \\ 0 \\ u_{3} \end{pmatrix}, \ 
\begin{pmatrix} 0 \\ u_{3} \\ \infty \end{pmatrix}, \
\begin{pmatrix} u_{3} \\ \infty \\ -1/u_{3} \end{pmatrix}, \
\begin{pmatrix} \infty \\ -1/u_{3} \\ 0 \end{pmatrix}, \ 
\begin{pmatrix} -1/u_{3} \\ 0 \\ - u_{1} u_{3}^2 \end{pmatrix}, \cdots \Bigg\},
\label{map1p2qcyclic}
\end{equation}
and which was obtained, as explained in the introduction, by iterating the mapping for an initial condition with $u_2=\epsilon$ and then taking the limit $\epsilon\to0$. In analogy to the terminology we introduced in \cite{express} for the case of second order mappings, we shall refer to such a pattern as a {\sl cyclic} pattern, here with length 4.
The second singularity, at $X_{0}$ with $u_3=0$, corresponds to a standard, non-cyclic, confined singularity pattern
\begin{equation}
\Bigg\{ \begin{pmatrix} u_{1} \\ u_{2} \\ 0 \end{pmatrix}, \ 
\begin{pmatrix} u_{2} \\ 0 \\ \infty \end{pmatrix}, \
\begin{pmatrix} 0 \\ \infty \\ \infty \end{pmatrix}, \ 
\begin{pmatrix} \infty \\ \infty \\ 0 \end{pmatrix}, \ 
\begin{pmatrix} \infty \\ 0 \\ u_{2} \end{pmatrix} \Bigg\}.
\label{map1p2qopen}
\end{equation}
Note that in this singularity pattern we only show the initial space $X_0$ and its iterates up to the last one to have co-dimension greater than 1. For example, at the limit $\epsilon\to0$, the next iterate would be $^t\!\left(0~\,u_2~\,u_1\right)$ at which point the iterates regain the dependence on $u_1$ that was lost upon entering the singularity. As in \cite{intcriterion}, to differentiate this case from the cyclic one, we shall refer to such a pattern as an {\sl open} singularity pattern, in this case of length 5.

For $q \in \mathbb{Z}_{\geq3}$, the singularity pattern for $X_{0}$ with $u_{q+1}=0$ is obtained from the following iteration of the mapping:
\begin{equation}
\begin{aligned}
\mathbf{x}_{0} &= \begin{pmatrix} u_{1}  \\ \vdots \\ u_{q-1} \\ u_{q} \\ 0_1 \end{pmatrix}
\xmapsto{1}
\begin{pmatrix}  u_{2} \\ \vdots \\ u_{q} \\ 0_1 \\ \infty_1 \end{pmatrix} 
\xmapsto{2}
\begin{pmatrix} u_{3} \\ \vdots \\ u_{q} \\ 0_1 \\ \infty_1 \\ * \end{pmatrix} \mapsto \cdots \xmapsto{q-1}
\begin{pmatrix} u_{q} \\ 0_1 \\ \infty_1 \\ * \\  \vdots \\ * \end{pmatrix} 
\xmapsto{q}
\begin{pmatrix} 0_1 \\ \infty_1 \\ * \\  \vdots \\ * \\ \infty_2 \end{pmatrix} 
\xmapsto{q+1} 
\begin{pmatrix} \infty_1 \\ * \\  \vdots \\ * \\ \infty_2 \\ 0_2 \end{pmatrix} 
\xmapsto{q+2}
\begin{pmatrix} * \\  \vdots \\ * \\ \infty_2 \\ 0_2 \\ * \end{pmatrix} 
\\ & \mapsto \cdots \xmapsto{2q}
\begin{pmatrix} \infty_2 \\ 0_2 \\ * \\ \vdots \\ * \end{pmatrix} \xmapsto{2q+1}
\begin{pmatrix}  0_2 \\ * \\ \vdots \\ * \\ * \end{pmatrix},
\end{aligned}
\label{map1pqopen}
\end{equation}
where * indicates a regular value; the symbols $0_1, \infty_1, \infty_2$ and $0_2$ will be explained below. In this case, the co-dimension of each $X_j$ is 
$$
{\rm cod}(X_j)=
\begin{cases}
1, ~\ j=0 \ \text{and} \ j \geq 2q+1\\
2, ~\ 1 \leq j \leq q-1 \ \text{and} \ q+2 \leq j \leq 2q\\
3, ~\ j=q, q+1 .
\end{cases}
$$

In other words, the co-dimension $1$ subspace $X_0$ collapses to a co-dimension $2$ subspace, $X_1$, which, after another $q-1$ iterations of the mapping, enters a co-dimension $3$ subspace, $X_q$. After an additional $q+1$ iterations of the mapping, $X_q$ eventually again gives rise to a co-dimension 1 subspace, $X_{2q+1}$. That the singularity pattern confines is thanks to two loci of indeterminacy for $\fie_q$, at $X_{q+1}$ and $X_{2q}$. The detailed calculation goes as follows. Let us denote the 
$(q+1)$-st component in vectors $\mathbf{x}_0, \mathbf{x}_1,\mathbf{x}_{q}$ and $\mathbf{x}_{q+1}$ in the pattern \eqref{map1pqopen} as $0_1, \infty_1, \infty_2$ and $0_2$ respectively. These take the precise forms:
$$
\begin{aligned}
X_0:\quad 0_1 &:= \epsilon, \\
X_1:\quad \infty_1 &:= u_1 + \frac{1}{0_1} - \frac{1}{u_2} 
    = \epsilon^{-1} + u_1 - \frac{1}{u_2}, \\
X_q:\quad \infty_{2} &:= u_{q} + \frac{1}{f} -\frac{1}{0_1} 
    = -\epsilon^{-1} + u_{q} + \frac{1}{f}, \\
X_{q+1}:\quad 0_2 &:= 0_1 + \frac{1}{\infty_2} - \frac{1}{\infty_1}
    = -\epsilon - \epsilon^{2}\left(u_{q} + \frac{1}{u_2} + \frac{1}{f} - u_1\right) + \mathcal{O}\left(\epsilon^{3}\right),
\end{aligned}
$$
where $f$ is shorthand for the regular (generic) value that appears in the $(q+1)$-st component of vector $\mathbf{x}_{q-1}$:
\begin{equation*}
f= -\frac{1}{u_q} + \dfrac{u_2\big(\prod_{j=3}^{q-1} u_j^2\big)}{\sum_{k=2}^{q-2} \big((-1)^k u_2 \big(\prod_{j=3}^{q-k} u_j^2\big) u_{q-k+1}\big) + (-1)^{q-1}}.\label{fab1}
\end{equation*}

The $(q+1)$-st component of $\mathbf{x}_{q+2}$ is then given by
$$X_{q+2}:\quad \infty_{1} + \frac{1}{0_2} - \frac{1}{g}
    =  u_q + \frac{1}{f} - \frac{1}{g} + \mathcal{O}\left(\epsilon \right),
$$
which is regular due to an exact cancellation of the infinities; here $g$ represents the regular (generic) value of the second component in $\mathbf{x}_{q+1}$: $g=u_2-\frac{1}{u_3}$.  Note that due to the form of $f$,
$$f: -\frac{1}{u_q}+ \text{a non-zero rational function independent of } u_q,$$
the combination $u_q + \frac{1}{f} - \frac{1}{g}$ will always depend on $u_q$. Hence, the dependence on $u_q$ that disappeared in $\mathbf{x}_{q}$ has now reappeared in  the $(q+1)$-st component of $\mathbf{x}_{q+2}$ at $\epsilon\to0$ and $X_{q+2}$ therefore has co-dimension 2 since, at $\epsilon\to0$, $\mathbf{x}_{q+2}$ is still independent of $u_1$. 

Continuing the iteration, we find that the $(q+1)$-st component in $\mathbf{x}_{2q+1}$ is 
$$X_{2q+1}:\quad \infty_2+\frac{1}{h} -\frac{1}{0_2} = u_1 + \frac{1}{h} - \frac{1}{u_2}+ \mathcal{O}\left(\epsilon \right),$$
where $h$ stands for the regular (generic) value in the $(q+1)$-st  component  in $\mathbf{x}_{2q}$ which  does not depend on $u_1$. The $(q+1)$-st component in $\mathbf{x}_{2q+1}$ then, at the limit $\epsilon\to0$, does depend on $u_1$ (the dependence on which had disappeared after the very first iteration of the mapping)  and $X_{2q+1}$ has co-dimension 1 and the singularity has confined. 

Note that $X_{2q+1}\neq X_0$ and that, for generic values of $u_1, \cdots, u_q$, the iterates of vectors in this subspace by $\fie_q$ are expressed as rational functions of the original (generic) coordinates in $X_{2q+1}$ with ever increasing degrees. These spaces $X_j$ for $j\geq2q+1$ are therefore always of co-dimension 1 and the singularity pattern for $X_0$ with $u_{q+1}=0$ cannot be a cyclic one. Summarizing, we can conclude that the singularity pattern for $X_{0}$ with $u_{q+1}=0$ is an open one, with length $2q+1$, not only for $q=2$ but for any $q \in \mathbb{Z}_{\geq2}$. 

For $q \in \mathbb{Z}_{\geq3}$, the singularity pattern for $X_{0}$ with $u_2=0$ is obtained from the following iteration of the mapping:
\begin{equation}
\begin{aligned}
\mathbf{x}_{0}  = & \begin{pmatrix} u_{1} \\ 0 \\ u_{3} \\ \vdots \\ u_{q+1} \end{pmatrix}
\xmapsto{1}
\begin{pmatrix} 0 \\ u_{3} \\ \vdots \\ u_{q+1} \\ \infty \end{pmatrix} 
\xmapsto{2}
\begin{pmatrix} u_{3} \\ \vdots \\ u_{q+1} \\ \infty \\ -1/u_{3} \end{pmatrix} \xmapsto{3}
\begin{pmatrix} u_{4} \\ \vdots \\ u_{q+1} \\ \infty \\ -1/u_{3} \\ -1/u_{4} \end{pmatrix} \mapsto \cdots \xmapsto{q}
\begin{pmatrix} u_{q+1} \\ \infty \\ -1/u_{3} \\  \vdots \\ -1/u_{q+1} \end{pmatrix} \\ 
\xmapsto{q+1} &
\begin{pmatrix} \infty \\ -1/u_{3} \\  \vdots \\ -1/u_{q+1} \\ 0 \end{pmatrix} \xmapsto{q+2}
\begin{pmatrix} -1/u_{3} \\ \vdots \\ -1/u_{q+1} \\ 0 \\ * \end{pmatrix} \xmapsto{q+3}
\begin{pmatrix} -1/u_{4} \\ \vdots \\ -1/u_{q+1} \\ 0 \\ * \\ * \end{pmatrix} \mapsto \cdots \xmapsto{2q}
\begin{pmatrix} -1/u_{q+1} \\ 0 \\ u_{3} \\ * \\ \vdots \\ * \end{pmatrix} \mapsto \cdots .
\end{aligned}
\label{map1pcyclic}
\end{equation}

Note that the co-dimension of $X_{j}$ is 2 for $j=1,..., q+1$ (and that the $0$ and $\infty$ entries in $\mathbf{x}_{q+1}$ are reciprocal, up to a factor -1 when expressed in terms of $\epsilon$). As a result, the co-dimension of $X_{q+2}$ is again 1, and one might be led to believe that this singularity pattern is also an open one. However, as $X_{q+2}$ clearly lies in the backward orbit of $X_0$ (with $u_2=0$), this singularity pattern is of course a cyclic one: iterating beyond $X_{q+2}$ we obtain a sequence of co-dimension 1 subspaces $(X_{j})_{q+2 \leq j \leq 2q}$, where $X_{2q}=X_{0}$ (in fact, $X_{2q+k}=X_{k}$ for all $k \in \mathbb{Z}$). We can therefore conclude that the singularity pattern for $X_{0}$ with $u_{2}=0$  is cyclic, with length $2q$ for any $q \in \mathbb{Z}_{\geq2}$. 

\subsection{The non-integrable case $(a \neq b)$ of mapping \eqref{map1p} for $q\geq2$}
As in the integrable case, we may assume that $a=1$ and $b \neq 1$ by taking an appropriate scaling transformation. 

Let us first consider the case for $q=2$. 
The first forward iterates of the singularity $X_0$ with $u_{3}=0$ are given below:
\begin{equation}
\mathbf{x}_{0}=\hskip-1mm \begin{pmatrix} u_{1} \\ u_{2} \\ 0 \end{pmatrix}\! \xmapsto{1}
\begin{pmatrix} u_{2} \\ 0 \\ \infty \end{pmatrix}\! \xmapsto{2}
\begin{pmatrix} 0 \\ \infty \\ \infty \end{pmatrix}\! \xmapsto{3}
\begin{pmatrix} \infty \\ \infty \\ 0 \end{pmatrix}\! \xmapsto{4} 
\begin{pmatrix} \infty \\ 0 \\ \infty \end{pmatrix}\! \xmapsto{5} 
\begin{pmatrix} 0 \\ \infty \\ \infty \end{pmatrix}\! \xmapsto{6}
\begin{pmatrix} \infty \\ \infty \\ 0 \end{pmatrix}\! \xmapsto{7}
\begin{pmatrix} \infty \\ 0 \\ \infty \end{pmatrix}\! \xmapsto{8}
\begin{pmatrix} 0 \\ \infty \\ \infty \end{pmatrix}\! \mapsto \cdots.
\label{nmap1p2qopen}
\end{equation}
We see that ${\rm cod}(X_1)=2$ and ${\rm cod}(X_j)=3$ for $j \in \mathbb{Z}_{\geq2}$, and that all spaces $X_{2+3k}$ are equal for any $k\in \mathbb{Z}_{\geq0}$. The backward iterates
\begin{equation}
\mathbf{x}_{0}= \begin{pmatrix} u_{1} \\ u_{2} \\ 0 \end{pmatrix} \xmapsto{-1}
\begin{pmatrix} \frac{b}{u_{1}}-\frac{1}{u_{2}} \\ u_{1} \\ u_{2} \end{pmatrix} \xmapsto{-2}
\begin{pmatrix} * \\ \frac{b}{u_{1}}-\frac{1}{u_{2}} \\ u_{1} \end{pmatrix} \mapsto \cdots ,
\label{nmap1p2qopenback}
\end{equation}
where $*=\frac{(u_1u_2-1)(u_1-b u_2)-b u_1^2 u_2}{u_1(u_1-b u_2)}$, only have regular entries and for generic $u_1$ and $u_2$ we therefore have ${\rm cod}(X_j)=1$ for $j \in \mathbb{Z}_{\leq0}$. Thus, the mapping \eqref{map1p} for $q=2$ has an unconfined singularity pattern with a repeating sub-pattern of length $3$:
$$
\Bigg\{\begin{pmatrix} u_{1} \\ u_{2} \\ 0 \end{pmatrix},
\begin{pmatrix} u_{2} \\ 0 \\ \infty \end{pmatrix},
\begin{pmatrix} 0 \\ \infty \\ \infty \end{pmatrix},
\begin{pmatrix} \infty \\ \infty \\ 0 \end{pmatrix},
\begin{pmatrix} \infty \\ 0 \\ \infty \end{pmatrix},
\begin{pmatrix} 0 \\ \infty \\ \infty \end{pmatrix},
\begin{pmatrix} \infty \\ \infty \\ 0 \end{pmatrix}, 
\begin{pmatrix} \infty \\ 0 \\ \infty \end{pmatrix}, \cdots \Bigg\}.
$$

For general $q \in \mathbb{Z}_{\geq3}$, the singularity pattern for $X_{0}$ with $u_{q+1}=0$ corresponds to the following chain with an infinitely repeating sub-pattern of length $q+1$:
\begin{equation}
\begin{aligned}
& \mathbf{x}_{0}= \begin{pmatrix} u_{1} \\ u_{2} \\ \vdots \\ u_{q-1} \\ u_{q} \\ 0_1 \end{pmatrix}
\xmapsto{1}
\begin{pmatrix}  u_{2} \\ u_{3} \\ \vdots \\ u_{q} \\ 0_1 \\ \infty_1 \end{pmatrix} 
\xmapsto{2}
\begin{pmatrix} u_{3} \\ \vdots \\ u_{q} \\ 0_1 \\ \infty_1 \\ * \end{pmatrix} \mapsto \cdots \xmapsto{q-1}
\begin{pmatrix} u_{q} \\ 0_1 \\ \infty_1 \\ * \\  \vdots \\ * \end{pmatrix} 
\xmapsto{q}
\begin{pmatrix} 0_1 \\ \infty_1 \\ * \\ \vdots \\ * \\ \infty_2 \end{pmatrix} 
\xmapsto{q+1} 
\begin{pmatrix} \infty_1 \\ * \\ \vdots \\ * \\ \infty_2 \\ 0_2 \end{pmatrix} \xmapsto{q+2}
\begin{pmatrix} * \\ \vdots \\ * \\ \infty_2 \\ 0_2 \\ \infty_3 \end{pmatrix}  \\
&
\mapsto \cdots \xmapsto{2q}
\begin{pmatrix} \infty_2 \\ 0_2 \\ \infty_3 \\ * \\ \vdots \\ * \end{pmatrix} \xmapsto{2q+1}
\begin{pmatrix}  0_2 \\ \infty_3 \\ * \\ \vdots \\ * \\ \infty_4 \end{pmatrix} \mapsto \cdots .
\end{aligned}
\label{nmap1pqopen}
\end{equation}
As in the integrable case we have ${\rm cod}(X_j)=2$ for $1\leq j\leq q-1$ and ${\rm cod}(X_j)=3$ for $j=q, q+1$, for any (generic) value of $b\neq0,1$.
However, whereas in the integrable case we had a fortunate cancellation of infinities in the image of $X_{q+1}$ which made that ${\rm cod}(X_{q+2})$ became 2 instead of 3, no such cancellation happens in the present case and we still have ${\rm cod}(X_{q+2})=3$.
As before, let us denote the $(q+1)$-st component in vectors $\mathbf{x}_0, \mathbf{x}_1,\mathbf{x}_{q}$ and $\mathbf{x}_{q+1}$ in the pattern \eqref{nmap1pqopen} as $0_1, \infty_1, \infty_2$ and $0_2$. These take the precise forms:
$$
\begin{aligned}
X_0:\quad 0_1 &:= \epsilon, \\
X_1:\quad \infty_1 &= u_1 + \frac{1}{0_1} - \frac{b}{u_2} 
    = \epsilon^{-1} + u_1 - \frac{b}{u_2}, \\
X_q:\quad \infty_{2} &= u_{q} + \frac{1}{f} -\frac{b}{0_1} 
    = -b\epsilon^{-1} + u_{q} + \frac{1}{f}, \\
X_{q+1}:\quad 0_2 &= 0_1 + \frac{1}{\infty_2} - \frac{b}{\infty_1}
    = -\delta \epsilon + \mathcal{O}\left(\epsilon^{2}\right),
\end{aligned}
$$
where $f$ denotes the $(q+1)$-st component of $\mathbf{x}_{q-1}$ (which takes a regular value) and $\delta=b+\frac{1}{b}-1\neq1$. 
The $(q+1)$-st component of $\mathbf{x}_{q+2}$ is now given by
$$X_{q+2}:\quad \infty_3 = \infty_{1} + \frac{1}{0_2} - \frac{b}{g}
    =  \left(1-\frac{1}{\delta} \right) \epsilon^{-1} + \mathcal{O}\left(1\right),
$$
where $g=u_2-\frac{b}{u_3}$, and no cancellation of infinities occurs. Exactly the same thing happens at the next locus of indeterminacy for $\fie_q$, $X_{2q}$:
$$X_{2q+1}:\quad \infty_4 = \infty_{2} + \frac{1}{h} - \frac{b}{0_2}
    =  -b \left(1-\frac{1}{\delta} \right) \epsilon^{-1} + \mathcal{O}\left(1\right),
$$
where $h$ represents the $(q+1)$-st component  of $\mathbf{x}_{2q}$ (which takes a regular, generic, value).
There are no cancellations of infinities at subsequent steps either, resulting in a chain of spaces $X_j$ in which, starting from $j=q$, the same space appears every $(q+1)$-st step. 

It is easy to check that $X_0$ with $u_{q+1}=0$ is not a singularity of the inverse mapping \eqref{invmap1p}. Moreover, as for the case when $q=2$, when a vector in $X_0$ with $u_{q+1}=0$ is iterated backwards, the first entry of each $\mathbf{x}_{j} \ (j=-1,-2,...)$ will be regular, indicating that ${\rm cod}(X_j)=1$ for $j \in \mathbb{Z}_{\leq0}$ for all $q\geq2$. Since the $X_{q+k(q+1)}$ are the same for all $k\in \mathbb{Z}_{\geq0}$, by combining the results of \eqref{nmap1p2qopen} and \eqref{nmap1pqopen}, we conclude that when $a\neq b$, the mapping \eqref{map1p} has an unconfined singularity pattern with a repeating sub-pattern of length $q+1$, for any $q \in \mathbb{Z}_{\geq2}$.

The singularity that arises at $X_0$ with $u_2=0$ however is yet of a different type. When $q=2$, we find
\begin{equation}
\mathbf{x}_{0}=\hskip-1mm \begin{pmatrix} u_{1} \\ 0 \\ u_3 \end{pmatrix}\!\! \xmapsto{1}\!\!
\begin{pmatrix} 0 \\ u_3\\ \infty \end{pmatrix}\!\! \xmapsto{2}\!\!
\begin{pmatrix} u_3 \\ \infty \\ -\frac{b}{u_3} \end{pmatrix}\!\! \xmapsto{3}\!\!
\begin{pmatrix} \infty \\ -\frac{b}{u_3} \\ \frac{(b-1) u_3}{b} \end{pmatrix}\!\! \xmapsto{4}\!\!
\begin{pmatrix} -\frac{b}{u_3} \\ \frac{(b-1) u_3}{b} \\ \infty \end{pmatrix}\!\! \xmapsto{5}\!\!
\begin{pmatrix} \frac{(b-1) u_3}{b} \\ \infty \\-\frac{b(2b-1)}{u_3(b-1)} \end{pmatrix}\!\! \xmapsto{6}\!\!
\begin{pmatrix} \infty \\ -\frac{b(2b-1)}{u_3(b-1)} \\ \frac{2 u_3(b-1)^2}{b (2b-1))} \end{pmatrix}\!\!  \mapsto \cdots,
\end{equation}
i.e. a chain of spaces with a subpattern of spaces of co-dimension 2 that repeats with period 3, starting from $X_2$. If we now look at the backward evolution,
\begin{equation}
\mathbf{x}_{0}=\hskip-1mm \begin{pmatrix} u_{1} \\ 0 \\ u_3 \end{pmatrix}\!\! \xmapsto{-1}\!\!
\begin{pmatrix} \infty \\ u_1\\ 0 \end{pmatrix}\!\! \xmapsto{-2}\!\!
\begin{pmatrix} -\frac{1}{u1} \\ \infty \\ u_1 \end{pmatrix}\!\! \xmapsto{-3}\!\!
\begin{pmatrix} u_1(1-b)\\ -\frac{1}{u1} \\ \infty \end{pmatrix}\!\! \xmapsto{-4}\!\!
\begin{pmatrix} \infty \\ u_1(1-b) \\ -\frac{1}{u1} \end{pmatrix}\!\! \xmapsto{-5}\!\!
\begin{pmatrix} \frac{b-2}{u_1(1-b)} \\ \infty \\ u_1(1-b) \end{pmatrix}\!\! 
\!\!  \mapsto \cdots,
\end{equation}
we see exactly the same phenomenon. Hence, this singularity pattern is not a standard unconfined one, but rather an {\sl anticonfined} one, i.e. a pattern in which we have unconfined patterns for both the forward and the backward iteration of the mapping \cite{anticonf}. In general, for $q\geq2$, we have for the forward iteration of $X_0$ with $u_2=0$,
\begin{equation}
\mathbf{x}_{0}=\hskip-1mm \begin{pmatrix} u_{1} \\ 0 \\ u_3  \\ \vdots \\ u_{q+1}\end{pmatrix}\!\! \xmapsto{1}\!\!
\begin{pmatrix} 0 \\ u_3\\ \vdots \\ u_{q+1} \\ \infty \end{pmatrix}\!\! \xmapsto{2}\!\!
\begin{pmatrix} u_3\\ \vdots \\ u_{q+1} \\ \infty \\ -\frac{b}{u_3} \end{pmatrix}\!\! \mapsto \cdots
\xmapsto{q+1}\!\! \begin{pmatrix} \infty \\ -\frac{b}{u_3} \\ *\\ \vdots\\ * \end{pmatrix}\!\! \xmapsto{q+2}\!\!
\begin{pmatrix} -\frac{b}{u_3} \\ *\\ \vdots\\ *\\ \infty \end{pmatrix}\!\! \xmapsto{q+3}\!\!
\begin{pmatrix} *\\ \vdots\\ *\\ \infty \\ * \end{pmatrix}\!\! \mapsto \cdots ,
\label{forwardanti}
\end{equation}
i.e. a chain of co-dimension 2 spaces $X_j$, $j\geq1$, that appear with period $q+1$ starting from $X_2$.  The backward evolution is slightly more involved ($q\geq 3$):
\begin{equation}
\begin{aligned}
&\mathbf{x}_{0}=\hskip-1mm \begin{pmatrix} u_{1} \\ 0 \\ u_3  \\ u_4 \\ \vdots \\ u_q \\ u_{q+1}\end{pmatrix}\!\! \xmapsto{-1}\!\!
\begin{pmatrix} f\\ u_1 \\ 0 \\ u_3  \\ u_4 \\ \vdots \\ u_q \end{pmatrix}\!\! \mapsto \cdots \xmapsto{-(q-2)}\!\!
\begin{pmatrix} *\\ \vdots \\ * \\f \\ u_1\\ 0 \\ u_3 \end{pmatrix}\!\! \xmapsto{-(q-1)}\!\! 
\begin{pmatrix} \infty \\ * \\ \vdots \\ * \\ f \\ u_1 \\ 0\end{pmatrix}\!\! \xmapsto{-q}\!\!
\begin{pmatrix} *\\ \infty\\ * \\ \vdots \\ * \\ f \\ u_1 \end{pmatrix}\!\! \mapsto \cdots \xmapsto{-(2q-1)}
\begin{pmatrix} *\\ \vdots \\ \vdots \\ \vdots \\ * \\ \infty \end{pmatrix} \\[-4mm] \\ 
&
\xmapsto{-2q}\!\!
\begin{pmatrix} \infty\\ *\\ * \\ \vdots \\ \vdots \\ * \end{pmatrix}\!\!\xmapsto{-(2q+1)}\!\!
\begin{pmatrix} * \\\infty\\ *\\ \vdots \\ \vdots \\ * \end{pmatrix}\!\!\mapsto \cdots ,
\end{aligned}
\label{backanti}
\end{equation}
where $f=u_{q+1}+\frac{b}{u_1}-\frac{1}{u_q}$. Here all spaces $X_j$ with $0\leq j\leq 2-q$ have co-dimension 1, but all subsequent spaces have co-dimension 2 and there is a pattern repeating  with period $q+1$ that starts at $X_{-q}$. Summarizing, we can say that the singularity at $X_0$ with $u_2=0$ is anticonfining when $a\neq b$. The corresponding singularity pattern has subpatterns that repeat with period $q+1$, starting at $X_2$ for the forward iteration and at $X_{-q}$ in the backward direction.

\subsection{Degree growth: the express method}
We shall now use the express method we introduced in \cite{express}, to investigate whether this method which has been shown to yield the exact value of the dynamical degree for three point mappings \cite{intcriterion}, can also be applied successfully to the case of multi-point mappings such as those we are studying here. In particular, we shall try to use the express method  to obtain the dynamical degrees of the reduced mappings, for general $q\geq2$, for both the integrable ($a=b$) and non-integrable $(a\neq b)$ cases.

Let us first consider the case $a=b$. We have seen that in this case, each mapping in the family of reduced mappings \eqref{map1p} has an open and a cyclic singularity pattern. As is customary in the express method for second order mappings, we shall neglect the cyclic pattern and only consider the open one. Since the mapping \eqref{map1p} is only a rewriting of the \changes{$(q+2)$-point} mapping \eqref{equ1p}, and since the open singularity pattern corresponds to a singularity that is due to a special value for $x_q=u_{q+1}$ (here $u_{q+1}=0$), we can limit our analysis to the successive values $x_{q+1}, x_{q+2}, \hdots$ generated from the initial conditions by \eqref{equ1p}, i.e. to the $(q+1)$-st components of the vectors in the open singularity patterns \eqref{map1pqopen} and \eqref{map1p2qopen} for \eqref{map1p}:
\begin{equation}
0, \; \infty, \; f_1, \; \cdots, f_{q-2}, \; \infty, \; 0, \;  g_{1}, \; \cdots, g_{q-1},\label{fullopenpatt}
\end{equation}
where the $f_i$ and $g_i$ represent the $2q-3$ generic values among the $(q+1)$-st components of the vectors in \eqref{map1pqopen} or \eqref{map1p2qopen}.

Since we do not control the values of the $f_i$ or $g_i$, we shall concentrate on the occurrences of the values $0$ and $\infty$ in the pattern \eqref{fullopenpatt} and since none appear after the $(q+2)$-nd entry in \eqref{fullopenpatt}, we can limit our analysis to the `reduced' pattern
\begin{equation}
0, \; \infty, \; f_1, \; \cdots, f_{q-2}, \; \infty, \; 0.\label{openpatt}
\end{equation}

Let us denote the number of `spontaneous' appearances of $0$ at the $n$-th iterate of the mapping by $Z_n$. By spontaneous occurrences of the value 0 we mean occurrences that are not forced by the singularity pattern, but that are just the result of some special initial conditions $u_1, u_2 \hdots, u_q$. Using the above pattern \eqref{openpatt}, if we express the fact that the number of preimages of 0 under the $n$th iterate of $\fie_q$, which we denote by $d_{n}(0)$, must be equal to that for the value $\infty$, denoted $d_{n}(\infty)$, we obtain the expression
\begin{equation}
    Z_n + Z_{n-q-1} \simeq Z_{n-1} + Z_{n-q},
\end{equation}
where the $\simeq$ symbol indicates equality ``up to possible contributions of the neglected cyclic pattern".
From this relation we deduce the characteristic equation (by taking $Z_n \sim \lambda^n$)
\begin{equation}
    (\lambda^q - 1)(\lambda -1)=0,
\end{equation}
which clearly does not possess any roots that are greater than 1, for any value of $q\geq2$. In the express method this is taken as proof of the integrable character of the mapping \cite{express, intcriterion} (by integrability we mean here that the dynamical degree of the mapping is 1) and for second order mappings the absence of a real root greater than 1 is actually known to imply quadratic growth for the degree of the iterates of that mapping \cite{intcriterion}. For now there is no similar rigorous statement for higher order mappings, but it has been shown that \changes{all reductions of the discrete KP equation (and therefore also of the dKdV equation) of the type we consider, can} have at most quadratic growth \cite{maseint} and we can thus view the express method as having successfully detected integrability for the entire family of mappings \eqref{map1p} with $a=b$.

Of course, such an `indirect' verification cannot be deemed sufficient proof of the applicability of the express method to higher order mappings and, to reinforce our argument, we shall therefore use this method to calculate the exact dynamical degrees for all mappings in the family  \eqref{map1p} when $a\neq b$.

As we have shown, the reduced mapping \eqref{map1p} with $a \neq b$ has an unconfined singularity pattern and an anticonfined one.
From the analysis that lead to the anticonfined pattern (\ref{forwardanti},\ref{backanti}), it is clear that the multiplicities of the values $0$ and $\infty$ that appear in that pattern exhibit no growth at all: both values only appear as the result of a leading order $\epsilon^1$ or $1/\epsilon^1$ in the singularity analysis. Hence, as was done in \cite{unconfined} where we generalized the express method to equations with non-confining singularities, we shall neglect any contributions due to the anticonfined pattern just as we neglected contributions due to cycic patterns for the confining case.

We are then left with the unconfined patterns \eqref{nmap1pqopen} (for $q\geq3$) and \eqref{nmap1p2qopen} (for $q=2$), which both have a repeating subpattern of length $q+1$, starting at the $q$th iterate of the mapping, i.e. at the ($q+1$)-st entry in the pattern. As explained above, if we concentrate on the values $0$ and $\infty$ in these patterns, then it suffices to extract the last component of each vector in them, which yields the pattern:
$$
0, \; \infty, \; f_{1}, \; \cdots, \; f_{q-2}, \; \infty, \; 0, \; \infty, \; f_{1}^{'}, \; \cdots, \; f_{q-2}^{'}, \; \infty, \; 0, \; \infty,\;\cdots \,.
$$
As before, denoting the number of (spontaneous) appearances of the value $0$ by $Z_n$ and neglecting the contributions from the anticonfined pattern, the number of preimages of $0$ and $\infty$ under the $n$th iterate of $\fie_p$, due to this pattern, are given by
\begin{equation}
\begin{aligned}
d_n(0)\simeq Z_n + Z_{n-(q+1)} + Z_{n-2(q+1)} + \cdots &= \sum_{k=0}^{+\infty} Z_{n-k(q+1)}, \\
d_n(\infty)\simeq Z_{n-1}+Z_{n-q}+Z_{n-1-(q+1)}+Z_{n-q-(q+1)}+ \cdots &= \sum_{k=0}^{+\infty} Z_{n-1-k(q+1)} + \sum_{k=0}^{+\infty} Z_{n-q-k(q+1)},
\end{aligned}
\end{equation}
where we set $Z_{j<0}=0$. As explained in \cite{unconfined}, we can now extract a characteristic equation for $Z_j\sim\lambda^j$ from the relation $d_n(0)\simeq d_n(\infty)$, by assuming the existence of a characteristic root greater than 1 and taking the limit $n \rightarrow \infty$:
\begin{equation}
\frac{1}{1-\lambda^{q+1}}=\bigg(\frac{1}{\lambda}+\frac{1}{\lambda^{q}}\bigg)\frac{1}{1-\lambda^{q+1}}.
\end{equation}
This yields the characteristic equation
\begin{equation}
\lambda^{q}-\lambda^{q-1}-1=0.
\label{p1charequ}
\end{equation}
The largest root of \eqref{p1charequ} is greater than 1 for all $q \in \mathbb{Z}_{\geq2}$, indicating non-integrability for all the mappings  in the family \eqref{map1p}, when $a \neq b$. We have verified the values of the dynamical degrees numerically for all mappings in this family using Halburd's Diophantine method \cite{diophantine, detect}, in the range $2\leq q\leq10$, and found excellent agreement between the numerical values and the largest root of \eqref{p1charequ} for those values of $q$. This then indicates that the express method indeed correctly predicts non-integrability for all these mappings and, moreover, yields the exact value of their dynamical degrees.

When $q=2$, the largest root of \eqref{p1charequ} is equal to the golden mean $\fie=(1+\sqrt{5})/2 \approx 1.6180$. As a matter of fact, the largest root of \eqref{p1charequ} can be seen to be a Pisot number for $q=2,3,4,5$, i.e. an algebraic integer greater than 1, the Galois conjugates of which all lie inside the unit disc. However, when $q \in \mathbb{Z}_{>5}$ this largest root is not a Pisot number. This is easily seen from the fact that the largest root for  \eqref{p1charequ} decreases (strictly monotonically) in size with increasing $q$ and from the value the largest root takes at $q=5$. At that value the characteristic equation can be factored as $(\lambda^2-\lambda+1)(\lambda^3-\lambda-1)$ and the largest root comes from the cubic factor $(\lambda^3-\lambda-1)$ which tells us that it is, in fact, identical to the smallest Pisot number $\approx 1.3247...$ (a.k.a. the {\sl plastic number}). For $q>5$ the value of the largest root for \eqref{p1charequ} then descends below the plastic number and tends to 1 as $q\to\infty$. That the dynamical degree for mapping \eqref{map1p} (or equivalently, of equation \eqref{equ1p}) for $a\neq b$ must tend to 1 when $q\to+\infty$ is also readily seen on the equation itself: the first $q-1$ iterates $x_{q+j} ~(j=1,\hdots,q-1)$ of \eqref{equ1p} are all first degree rational expressions in the initial value $x_q=u_{q+1}$ (which is the only one that is of interest to the express method). Hence, when $q\to+\infty$, any degree growth with respect to this variable is postponed indefinitely and the dynamical degree is equal to 1. (When considered as rational functions of all initial values $u_1, \hdots, u_{q+1}$, the first $q-1$ iterates $x_{q+j}$ of \eqref{equ1p} have degree $2j+1$; this linear degree growth then persists indefinitely when $q\to+\infty$, leading to the same conclusion.)

\section{Singularity patterns for the discrete KdV equation}\label{KdVsing}
In this section, we will show that the study of the singularities for the reduced mapping \eqref{map1p} also allows us to discuss the properties of certain types of singularities that arise for the dKdV equation, not only for periodic intial conditions such as those required for the reduction $x_{m+1,n}=x_{m,n+q}$, but also for general initial conditions, on initial staircases with varying widths.

\subsection{An `open' singularity pattern for dKdV due to mapping  \eqref{map1p}}\label{sec31}
\begin{figure}[h]
\begin{center}
\resizebox{12.5cm}{!}{\includegraphics{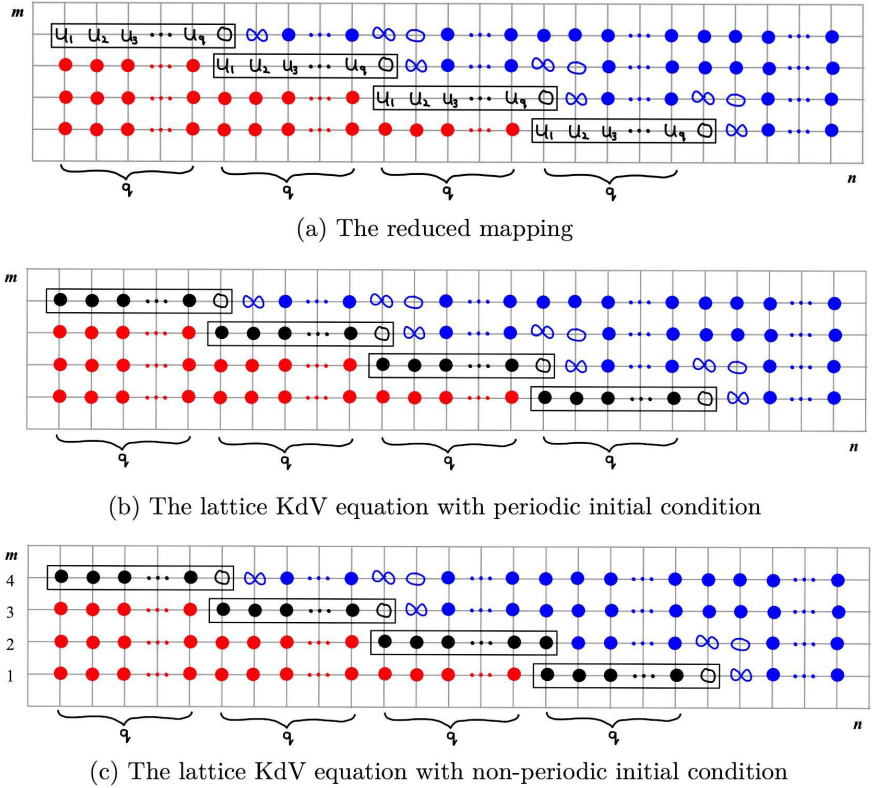}}
    \caption{Singularity patterns for dKdV obtained from mapping \eqref{map1p} for $X_0$ with $u_{q+1}=0$.}
\label{p1openUq1=0}
\end{center}
\end{figure}

Consider for example the open singularity patterns of the reduced mapping \eqref{map1p}, with $a=b$, for $q \geq 2$. The reduction condition $x_{m+1,n}=x_{m,n+q}$ establishes a (bijective) correspondence between the open chain \eqref{map1pqopen} in Section \ref{redmapp1} and the open singularity pattern on the lattice shown in Figure \ref{p1openUq1=0} (a), for the original dKdV equation \eqref{dKdV} with periodic initial conditions (on a staircase of height 1 and width $q$). This pattern remains unchanged if, instead of the purely periodic initial condition of Figure  \ref{p1openUq1=0} (a), we choose initial conditions such as in Figure \ref{p1openUq1=0} (b) where a $0$ still appears periodically but the other (non-zero) initial values are now no longer necessarily the same in each row. 
Going yet one step further, one may also choose to break the remaining periodicity in the initial condition by choosing only nonzero regular values in certain rows. In this case,  the local singularity patterns, the squares of infinities and zeros seen in Figures  \ref{p1openUq1=0} (a) and (b), will simply disappear in the rows that do not contain a $0$ and the remainder of the pattern is unchanged, as shown in Figure \ref{p1openUq1=0} (c). 

Note that, locally, the singularity pattern arising from a single 0 (at a corner on the staircase) is nothing but the pattern shown in Figure \ref{Fig2}, which is of course the basic singularity pattern if there is only one such zero. In section \ref{classification} we shall investigate what happens if there are (finitely) many such zeros on different steps of the initial staircase.

\begin{figure*}[h]
\begin{center}
\resizebox{12.5cm}{!}{\includegraphics{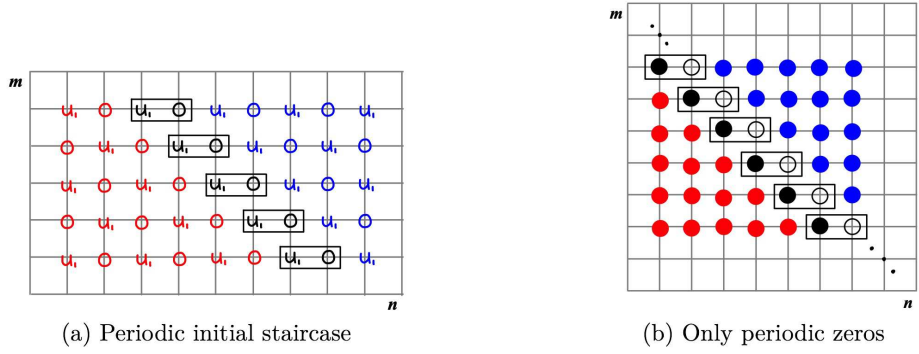}}
\caption{The singularity pattern on the lattice for $X_0$ with $u_{q+1}=0$ when $q=1$.}
\label{p1q1sing}
\end{center}
\end{figure*}

Note also that the case $q=1$ is special, since it corresponds to a 2-periodic reduced mapping without singularities but with infinitely many repeating 0 values:
$$
\left\{\ve{u_1}{0},\ve{0}{u_1},\ve{u_1}{0},\ve{0}{u_1},\ve{u_1}{0},\ve{0}{u_1},\cdots \right\}.
$$
The corresponding `singularity' pattern on the lattice is shown in Figure \ref{p1q1sing} (a). However, if as in Figure \ref{p1q1sing} (b) the non-zero values on the staircase are all different then no singularity appears. If only a few regular values on the staircase are different, then the zeros seen propagating along straight lines $m=n$ in Figure \ref{p1q1sing} (a) will disappear at some locations.

\subsection{An anticonfining singularity for dKdV}\label{anticonf}\label{sec32}
\begin{figure}[h!]
\centering
\resizebox{13.15cm}{!}{\rotatebox{90}{\includegraphics{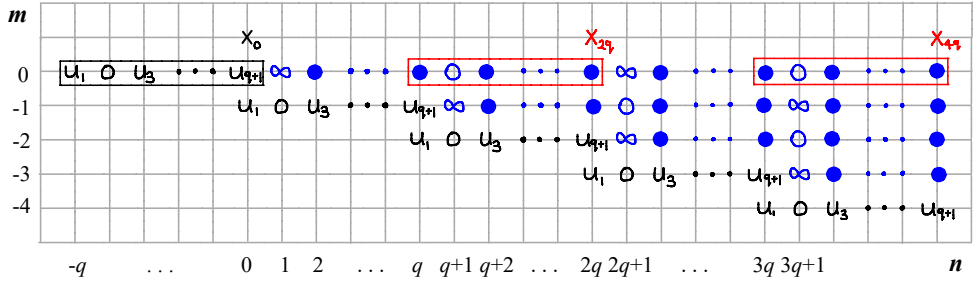}}}
\caption{The cyclic singularity pattern for $X_0$ with $u_2=0$ under the reduction $x_{m+1,n}=x_{m,n+q}$.}
\label{p1u2=0}
\end{figure}

We now consider the case of the cyclic singularity pattern for the reduced mapping \eqref{map1p}, with $a=b$, for $q \geq 2$.
Figure \ref{p1u2=0} shows two cycles in row $m=0$ of the cyclic chain \eqref{map1pcyclic}, on the lattice, when the $(q+1)$-st entry, $u_{q+1}$, of $\mathbf{x_0} \in X_0$ is fixed at $(0,0)$. By imposing periodicity as in the reduction $x_{m+1,n}=x_{m,n+q}$, we obtain the values in rows $m=-1,-2,-3,-4, \cdots$. This pattern indeed corresponds to the singularity pattern one obtaines for the dKdV equation \eqref{dKdV} when the initial data is given on the staircase shown in Figure \ref{p1u2=0}. In fact, it is easily verified that even for non-periodic non-zero values on such a staircase, with the same positions of the value $0$, one still has the same singularity pattern. In this pattern we now have lines of alternating zeros and infinities rising straight up, periodically, in the lattice. The pattern is therefore unconfined but, just as for the unconfined cases explained in the introduction, this can be put down to the fact that we have infinitely many singularities on the initial staircase.

\begin{figure}[h]
\begin{center}
\resizebox{6.cm}{!}{\includegraphics{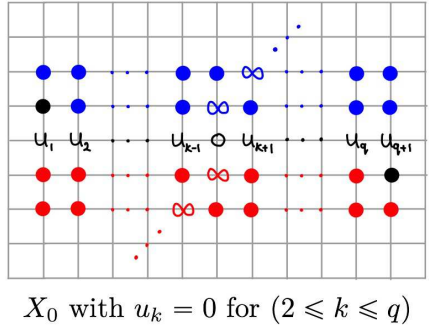}}
\caption{An anticonfined singularity pattern for dKdV.}
\label{Locsing}
\end{center}
\end{figure}

However, it is easily checked that a non-confining pattern in which  infinities emanate from a zero indefinitely along the line $m=n+1$ $(n \in \mathbb{Z}_{\geq n_0})$ already arises even for a {\sl single} singularity on the initial staircase, if this singularity corresponds to  $X_0$ with $u_2=0$ for the reduced mapping \eqref{map1p}. In fact, as depicted in Figure \ref{Locsing} (a), any initial value $u_k=0$ for {\sl any} $k$, $2 \leq k \leq q$,  leads to such a singularity pattern for the dKdV equation. 

A small remark is in order here. It might seem that for this type of singularity there is no one-to-one correspondence between the singular behaviour of the reduced mapping and that of dKdV (with initial values on a staircase), but this is not the case. For subspaces $X_0$ \eqref{X0} with $u_k=0$ $(3 \leq k \leq q)$, the reduced mapping is not singular at the very first iteration, but it is singular at the $(k+1)$-st iteration (${\rm cod}(X_{k+1})=2$). I.e., since $u_k$ is the second entry in $\mathbf{x}_{k+1} \in X_{k+1}$, when it comes to the correspondence with singularities for dKdV, the $q-2$ subspaces $X_0$ with $u_k=0$ $(3 \leq k \leq q)$  for the reduced mapping all play exactly the same role as the subspace $X_0$ with $u_2=0$. Therefore, the singularity $X_0$ with $u_2=0$ for the reduced mapping is also representative of singularities such as in Figure \ref{Locsing}. 

The existence of a non-confining singularity might seem to contradict the integrability of the dKdV equation but this need not necessarily be the case. Up to now we have only considered evolutions of the dKdV equation in the northeast direction. However, if we calculate the backward evolution in the southwest direction for the initial conditions on the staircase in Figure  \ref{Locsing}, we see that the 0 that causes this singularity must also have infinities in its `past', extending indefinitely along the line $m=n-1$ $(n \in \mathbb{Z}_{\leq n_0})$. This singularity pattern is therefore similar in nature to the anticonfined patterns we encounter for ordinary mappings, and we shall call it {\sl anticonfined} for the dKdV equation as well. That such a singularity is not an impediment to integrability is because it is akin to a ``fixed" singularity for the equation: the zero that is at its origin cannot arise `spontaneously' through some accidental interaction of initial values. As we have seen, it can only arise as the consequence of a semi-infinite line of infinities as in Figure \ref{Locsing} or, as the result of infinitely many zeros along a line $m=n_0-n$ $(n \in \mathbb{Z}_{>0})$ as in Figure \ref{p1q1inf} (a). 

\subsection{Classification of singularity patterns for dKdV based on mapping \eqref{map1p}}\label{classification}
All this leads to the question: what can one say about general singularity patterns for the dKdV equation? As we mentioned in the introduction, in a general setting this seems to be an intractable problem, but using the results obtained in the previous sections we can narrow it down to a more manageable one for a subclass of singularity patterns for KdV: what are the singularity patterns for the dKdV equation for initial values given on a staircase with uniform height 1 and varying width $q_j$, that are associated with co-dimension 1 singularities for reduced mappings obtained from periodic repetitions of each width $q_j$ part of the staircase?
As explained in sections \ref{sec31} and \ref{sec32}, such singularities can only arise from a 0 at either a position $u_{q+1}$ on a step of the staircase (i.e. at a protruding corner of the staircase, cf. figure \ref{Redp1}) which, when isolated, corresponds to an open singularity pattern, or at a position $u_j$, for some $2\leq j\leq q$, corresponding to an anticonfining pattern for dKdV. 

Since our goal is to understand what can be learned from the singularity analysis for the family of reduced mappings \eqref{map1p} (with $a=b$), we shall not investigate possible interactions between singularity patterns that arise from singularities on the {\sl same} step of an initial staircase for dKdV, as these would correspond to singularities for the reduced mapping with codimension greater than 1. Hence, we shall only consider combinations of (finitely many) zeros, each located on a {\sl different} step of an initial staircase of height 1, with steps of varying width. Moreover, as should be clear from the analysis in sections \ref{sec31} and \ref{sec32}, when these singularities lie on non-adjacent steps on the initial staircase, their respective patterns never interact and the ensuing, global, singularity pattern is simply a superposition of the individual patterns. If we therefore limit ourselves to the case of singularities on adjacent steps on the initial staircase, this leaves five cases to consider.

\subsubsection*{Case 1: open + open (adjacent singularities)}
Let us first consider what happens when the initial staircase for the dKdV equation is such that two open singularity patterns on different parts of the staircase overlap. For staircases of height 1 the only possible case arises when, locally, the initial staircase is a combination of staircases of widths $q_1\in\mathbb{Z}_{\geq1}$ and $q_2=1$ (in descending order of the steps of the staircase) for which the singularities, at $X_0$ with $u_{q_1+1}=0$ and $u_{2}=0$ respectively, will overlap. This is a singular initial condition which is locally identical to that obtained from a 1-1 staircase,  and the resulting singularity pattern is a rhombus shaped one, as those shown in Figures \ref{Caseopop} (a) and (b) in the introduction. 

\subsubsection*{Case 2: open + open (separated singularities)}
If two open singularity patterns on different parts of the staircase do not arise from diagonally adjacent zeros, then both patterns are preserved unperturbed in the lattice. However, even when the singularities themselves are not adjacent, it is still possible to have contiguous singularity patterns: for a staircase with uniform height 1 this happens for singular initial conditions, located at $X_0$ with $u_{q_1+1}=0$ and $u_{3}=0$ for the reduced mappings, on neighbouring steps of staircases of widths $q_1\in\mathbb{Z}_{\geq1}$ and $q_2=2$ respectively.

\subsubsection*{Case 3: anticonfining + open (adjacent singularities)}
\begin{figure}[h]
    \centering
    \resizebox{4.15cm}{!}{\includegraphics{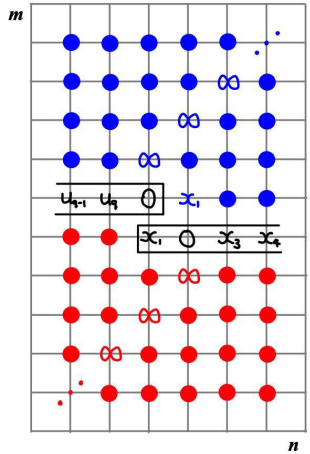}}
    \caption{Anticonfining + open (adjacent) singularity pattern on the lattice.}
    \label{Case5cycop}
\end{figure}

Next, let us consider the singularity pattern for the dKdV equation for an initial staircase where an open and an anticonfining singularity pattern overlap, as in the pattern shown in Figure \ref{Case5cycop}. 
For staircases with height 1,  the only possibility for two singular conditions corresponding to an anticonfining and an open singularity pattern for dKdV to be adjacent, is a combination of singularities of types $u_{q_1+1}=0$ and $u_{2}=0$ on neighbouring steps of the staircase of widths $q_1$ and $q_2$, respectively, where $q_1 \in \mathbb{Z}_{\geq 1}$ and $q_2 \in \mathbb{Z}_{\geq 2}$. In that case, the singularity pattern that arises from the open singularity disappears and the line of infinities that emanates from the anticonfining singularity now appears above the zero initial condition of the open singularity instead of the anticonfining singularity. 

\subsubsection*{Case 4: anticonfining + open (separated singularities)}
\begin{figure}[h]
\begin{center}
\resizebox{4.25cm}{!}{\includegraphics{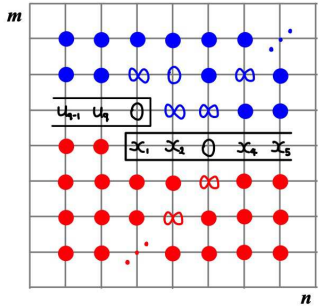}}
\caption{Anticonfining + open (separated) lattice singularity pattern.}
\label{Case6cycop}
\end{center}
\end{figure}

There are also singularity patterns for the dKdV equation for an initial staircase with an anticonfining and an open singular condition as in Figure \ref{Case6cycop}. If the two singularities are far enough apart then their respective patterns for the forward (as well as for the backward) iterations are preserved. However, these patterns can be contiguous on the lattice: this  happens for singularities of type $X_0$
with $u_{q_1+1}=0$ and $u_{3}=0$ for the reduced mapping, for adjoining staircases of respective widths $q_1$ and $q_2$, where $q_1 \in \mathbb{Z}_{\geq 1}$ and $q_2 \in \mathbb{Z}_{\geq 3}$. 

\subsubsection*{Case 5: anticonfining + anticonfining}
Finally, it is also possible to have combinations of only anticonfining patterns. As explained in section \ref{anticonf} such singularities can be located at any one of the initial conditions $u_2, \hdots, u_q$ of the corresponding reduced mapping, but whatever their precise position, they are always at least two lattice sites removed from the anticonfining singularity on the neighbouring step (on the right). Hence they are always separated and the lines of infinities (both in the forward as in the backward directions) that emanate from each singularity never interact.

\medskip
One can of course also consider different combinations of the above cases, but there will be no interactions between the respective local singularity patterns due to the fact that infinities always propagate along straight lines in the northeast direction and so are parallel to each other. The only possible interactions among singularities are described in Cases 1 and 3, where singularities have adjacent zeros: an open + open pattern yields an (extended) rhombus-like singularity pattern for dKdV (when there are only  finitely many singularities). In the case of an open + anticonfining pattern we observe a shifting role of the singularities, as the straight line of infinities of the cyclic pattern is passed on to the open one, as shown in Figure \ref{Case6cycop}.

\section{Conclusion}\label{conc}
In this paper we have studied the singularities of the dKdV equation from the viewpoint of the ARS method. More precisely: we have studied the co-dimension 1 singularities that arise in the mappings that are obtained as periodic reductions of the type $x_{m+1,n}=x_{m,n+q}$, of dKdV, for general $q\in\mathbb{Z}_{\geq1}$ and we have made systematic use of the results on the singularity structure of these mappings to describe a large class of singularity patterns for the dKdV equation. Among these patterns there are anticonfined ones, possibly combined with regular, confined, ones. To the best of our knowledge, the \changes{possibility} of having anticonfining singularities for the dKdV equation has not been reported before.

For the reduced mappings we also implemented a version of the express method we introduced in \cite{express} as a fast way to calculate dynamical degrees for second order mappings, and which is based on Halburd's original method for finding the exact degree growth for such mappings \cite{rodzero}. Here we have shown that the express method can be successfully applied to all mappings obtained in our reduction approach. The method not only successfully predicts integrability, it also yields the exact value of the dynamical degree for the reductions of a non-integrable extension of the dKdV equation. This is quite remarkable since the method was not designed to deal with higher order mappings.
 
We also have obtained partial results on the singularity structure for the  mappings that arise from the general reduction, $x_{m+p,n}=x_{m,n+q}$ for general $p,q\in\mathbb{Z}_{\geq1}$, which allow us to treat an even wider range of singularity patterns for the dKdV equation, with even more interesting interactions between them. The calculations for the general case, however, are obviously far more involved than those presented here and we shall report on them in a separate publication. 

\section*{Acknowledgements}
RW would like to acknowledge support from the Japan Society for the Promotion of Science (JSPS)  through JSPS grant number 18K03355. He would also like to thank Y. Ohta for pointing out the results of Z. Tsuboi in references \cite{gromov} and \cite{tsuboi}.


\end{document}